# Gold Nanoparticles supported on MoS$_2$ Nanoribbons matrix as a biocompatible and water dispersible platform for enhanced photothermal ablation of cancerous cells using harmless near infrared irradiation.

*Shobhit Pandey[a,#], Anushka Bansal[a,#], Vinod Kumar[b], Himanshu Mishra[c], Rohit Kumar Agrawal[a], Sunayana Kashyap[d], , Bratindranath Mukherjee[a], Preeti Saxena[d], and Anchal Srivastava[c*]*

[a] *Department of Metallurgical Engineering, Indian Institute of Technology (Banaras Hindu University), Varanasi-221005, India*

[b] *Institute of Nano Science and Technology, Habitat Centre, Phase- 10, Sector- 64, Mohali, - 160062 India*

[c] *Department of Physics, Banaras Hindu University (BHU), Varanasi-221005, India*

[d] *Department of Zoology, Banaras Hindu University, Varanasi, 221005, India*

*# Equally contributing authors.*

**Address correspondence to:**

E-mail: anchalbhu@gmail.com (Dr. AnchalSrivastava)

Phone No.: +91-9453203122, Fax: +91-542 2368174

## ABSTRACT

With photothermal efficiency higher than traditionally used reduced graphene oxide, while simultaneously being water dispersible, we propose biocompatible Molybdenum Nanoribbons - gold Nanoparticles (MoS$_2$ NR-Au NPs) system for synergistically enhanced photo-thermal ablation of cancerous cells (100% fatality) using NIR radiation (808nm), owing to commendable temperature rise above 70°C at a much faster rate.

## COMMUNICATION

The ubiquitous torment of Cancer has been prevalent for decades owing to the convoluted multi-level complexities worsened by the arcane knowledge of these deadly cells.[1,2] With the Nanotechnology at its apex, nanoparticles have been extensively employed for Cancer therapy. In recent times, Near Infrared Window (NIR) (700-1300nm) has been engendered for photothermal therapy, circumventing the drawbacks of conventional chemotherapy which includes high side effects, meagre solubility and inadequate specific drug delivery efficiency.[3,4] The biological tissues are effectively transparent in the NIR range attesting potential for Bio imaging and triggering drug release mechanism in both In-vivo and in-vitro.[3,5] In presence of an efficacious nanomaterial candidate, a suitable NIR energy (808nm for present work) is efficiently absorbed and transduced into heat energy resulting in cascading photo ablation of cancer cells which cannot withstand a temperature above 40-45°C.[5] 808 nm Laser has been chosen for the present work owing to its high spatial resolution, extricating itself from the disadvantages of preponderantly used 980 nm laser, i.e. minimum absorption by water, keeping the healthy tissues benign.[6]

Among the potential photothermal agents, anisotropic gold nanostructures produced by cumbersome synthesis routes had their head start, allowing their band gap to be tailored into the NIR region for commendable heat transduction efficiency, which is further exalted by good biocompatibility.[7–10] Nevertheless, these uneconomic noble local Plasmon receptors for NIR absorbance lose their efficacy by dissolution of its non-equilibrium shape into spheres on heating.[11,12]

Voracious investigations to find impeccable alternatives to gold soon found 2D layered materials as its next potential domain extolled for their startling high surface area to mass ratio. Down this line graphene[13], graphene oxide [14,15], reduced graphene oxide [16,17] and copper chalcogenides[11,18] have shown potential but their usage is limited by their hydrophobic nature requiring meticulous surface fabrication to enhance their dispersion, resulting in unavoidable adulteration in the absorption power.[19]

Owing to similar 2D layered structures like graphene, transition metal dichalcogenides (TMDs) have shown extensive potential in electronics, sensing, photonics, catalysis, bioengineering, energy harvesting, and flexible electronics, but have not been much exploited in Biomedicines.[20] Very recently water dispersible $MoS_2$ has emerged as an efficient photo thermal therapy candidate.[21] Mechanically exfoliated $MoS_2$ and chemically exfoliated $MoS_2$ Nano sheets and monolayers have been attested for enhanced photothermal Cancer therapy. [22–24] Further, reports have shown the synergistic effects of combing gold nanoparticles with $MoS_2$ showing increase in thermal conductivity attributing to the high thermal conductivity of gold islands which enhances phonon transport on $MoS_2$ platform.[25]

Motivated by the reclusive success of Gold nanoparticles and $MoS_2$ for transduction of NIR into heat, while realising the potential of synergistic effects of combining them together, the present work manifests facile synthesis of biocompatible and water dispersible $MoS_2$ NR-Au NPs as a potential photo-thermal agent, which to our knowledge has not been reported yet. When tested under 808nm NIR laser, $MoS_2$ NR-Au NPs show high synergistically enhanced photothermal response with temperature rising up to 70°C in less than 3 minutes, for concentrations of 300 ppm and above, without compromising biocompatibility; attesting it as a better photothermal agent in comparison to both; the recent reports of $MoS_2$ which required uneconomic synthesis routes and in comparison with traditionally used RGO based systems which were not water dispersible. [22–24] The proposed synthesis route, as shown schematically in Figure 1 a), typically involves $H_2S$ gas based reduction of hydrothermally synthesised $MoO_3$ nanorods to $MoS_2$NR, followed by microwave decoration of gold nanoparticles (Refer ESI), which unlike mechanical exfoliation of $MoS_2$ nanosheets, provides potential for bulk production.

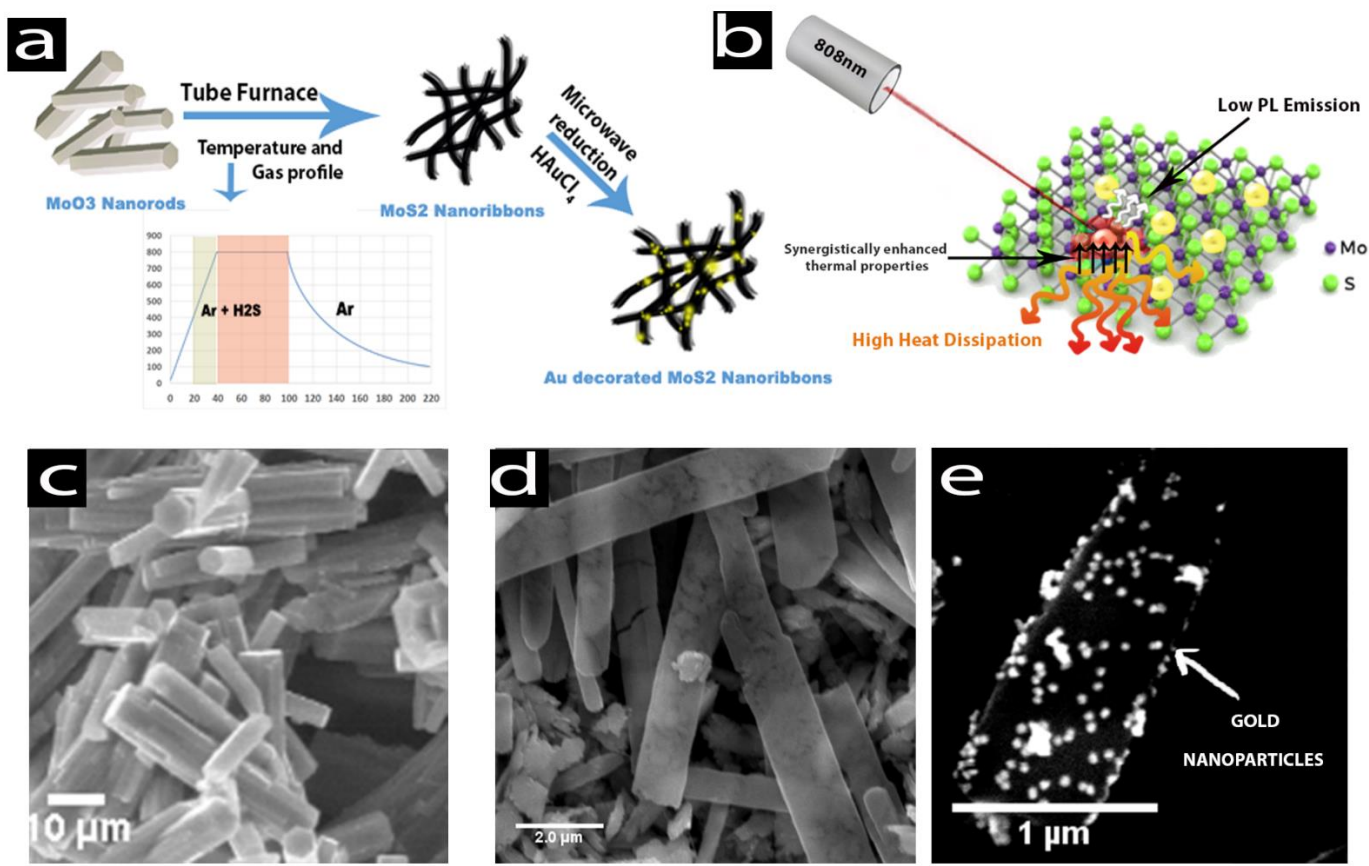

Figure 1: Synthesis and SEM characterizations. a) Schematic of Synthesis route adopted showing the intermediate and the final MoS$_2$ NR - Au NPs system. b) Synergistically enhanced heat dissupation under NIR radiation by MoS$_2$ NR- Au NPs system. c) SEM micrograph of MoO$_3$ nanorods d) SEM micrograph of MoS$_2$ nanoribbons d) SEM micrograph of MoS$_2$ NR- Au NPs system.

The size, morphology, and microstructures of the MoS$_2$ NR – Au NPs and the other intermediate products including MoS$_2$- NR and MoO$_3$- nr were determined by SEM (Hitachi S-4800). The SEM image in Fig 1 d) indicates that the final of yield of MoS$_2$ NR is composed of bundle-like nano ribbons of width 150 nm - 700 nm (large range due to the bundle like structure). These results are corroborated by the TEM micrographs of MoS$_2$ (Refer ESI) and the XRD results, were in the diffraction pattern when indexed, attests the identity of 2H – MoS$_2$ (Refer ESI), while the peaks when indexed in XRD further confirms the MoS$_2$ identity. Further the SEM investigation of the final product that is gold MoS$_2$ NR – Au NP system as seen in Figure 1 e) presents a strong evidence of nanoparticles of gold in the form of bright dots supported by the MoS$_2$ nano-ribbons.

For measuring the photothermal effect of the MoS$_2$ NR- Au NPs system NIR radiation from a 808 nm laser (output power 1W/cm$^{-2}$)was passed through a quartz cuvette containing an aqueous dispersion (200 µL) of MoS$_2$ NR – Au NPs with varying concentrations of from 10 to 600 ppm; one at a time. The dispersions were created by using simple sonication (20 min at 40KHz in water) of water containing as synthesised MoS$_2$ NR – Au NPs. The stability of the obtained dispersion was appreciable with no substantial sedimentation even after a week's time. Further the negligible effect of NIR on water, which ensures full photo transduction credibility to the hybrid system, was also attested in the same way by irradiating normal tap water, in which the dispersion was made. The quartz cuvette was kept at a distance of 5 cm from the laser (1 W/cm$^{-2}$). A digital thermometer equipped with a J-type thermocouple (accuracy of ±0.1 °C) was inserted into the aqueous dispersion a position where the probe of the thermocouple was not affected directly by the incoming radiation. The temperature was recorded at regular intervals of 10 sec, till the saturation temperature was reached. The results are shown in Figure 2. Evidently Figure 2 a) confirms the superior NIR photo thermal transduction by MoS$_2$ NR-Au NPs of concentration as low as 25ppm. Owing to this, rise in solution temperatures above the required threshold for thermal ablation of cancerous cells (40°C) in about 6 minutes was observed. Reference Much rapid transduction occurs at higher concentrations (above 300 ppm), with temperature reaching >40°C within a single minute and in 5 minutes crosses a mark of 70°C, which is much superior to the traditionally used GO[15], RGO [17] and even the recently reported chemically exfoliated MoS$_2$[26]. Further Fig 2 b) shows the maximum temperature rise that can be achieved as a function of the concentration of the MoS$_2$ NR – Au NPs system. The synergistic effect of combining MoS$_2$ with gold on enhanced thermal diffusivity has been shown by Sreeprasad.[25]. Also heat dissipation in photothermal is attributed to high absorption of the radiation, of which majority goes in heat dissipation (with low PL emission). Here combined photothermal effect from both Au NP and underlying MoS$_2$ NR matrix, which is further augmented by higher thermal diffusivity results in an overall higher heating effect. This is schematically shown in Fig 1 b).

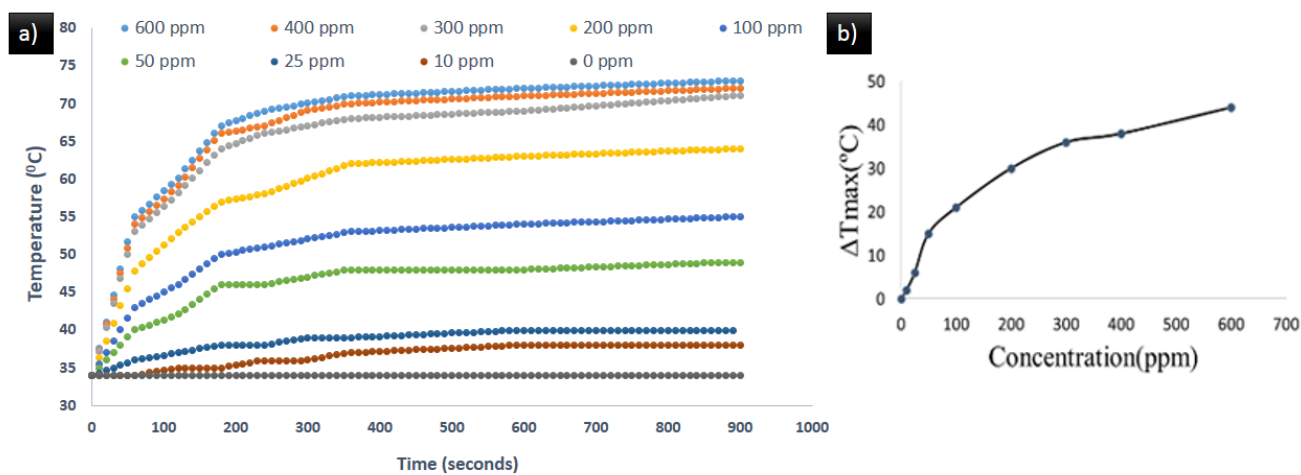

Figure 2: Photothermal Effect by MoS$_2$ NR- Au NPs system. a) Temperature vs Time profile for varying concentrations of MoS$_2$ NR- Au NPs on irradiation with NIR ( 808 nm) Laser. b) Max Temperature change saturation achieved for various concentrations of the system.

Though good photo transduction effect ensuring efficient conversion of electromagnetic energy to heat energy is the most important parameter but definitely it is not the only required parameter to prove the potential of a photo thermal agent to be brought on field. With this realization the toxicity tests were performed. The cytotoxicity result of MoS$_2$ NR-Au NPs have been shown in Fig 3 A). Evidently upto 300 ppm of MoS$_2$ NR-Au NPs, which itself has an ability to reach beyond 70$^0$ C, no significant toxicity have been estimated. At 400 ppm a marked change in cell viability is noticed.

With appreciable cytotoxicity results, the work next involves the NIR mediated phototherapy of SiHA cancerous cells. In vitro photothermal ablation effect of MoS$_2$ NR-Au NPs was examined in a detailed way. SiHa cancer cells were incubated with and without 200 ppm of MoS$_2$ NR-Au NPs were irradiated to 808 nm (1W/Cm$^2$) for 3 minutes. After irradiation cells were stained with trypan blue (0.4%) for 5 minutes to see the effect of laser treatment. Optical microscopic images of laser treated cells after staining with trypan blue have been shown in Fig. 3 C. Evidently, Fig. 3 C (a) represents the control healthy cells without laser irradiation in absence of MoS$_2$ NR- Au NPs. Cell viability 24 hrs post laser irradiation was estimated to probe the phtotothermal efficiency of MoS$_2$ NR-Au NPs (Fig. 3 B). Evidently in Fig 3 B), cells incubated with 200 ppm of MoS$_2$ NR-Au NPs without NIR exposure show comparable cell viability to that of control cells. However, in the same figure we can see that the cells incubated with 200 ppm of MoS$_2$ NR-Au NPs show only 10% of cell viability after NIR exposure in comparison to that of control cells. This is not surprising as solution temperatures reach well above thermal ablation thresholds. Cell viability results after NIR exposure clearly describes the tremendous photothermal effects of MoS$_2$ NR-Au NPs. This can further be proved by the optical micrographs of the laser irradiated SiHa cells as shown in Fig 3 C). Fig. 3 C (d) is the characterization image of cells cultured with 200ppm MoS$_2$ NR-Au NPs followed by laser irradiation. In trypan blue staining, nearly all cells have taken up blue stain [Fig. 3 C (d)] i.e. dead. This indicates the SiHa cancer cells were completely killed just in 3 minutes of NIR laser irradiation in presence of MoS$_2$ NR-Au NPs. Approximate 100 % killing of the cancer cells have been achieved MoS$_2$NR-AuNPs mediated phototherapy of SiHa cancer cells. This is in stark contrast to the cells treated with MoS$_2$ NR-Au NPs without irradiation. There, the cell viability was observed to be approximately 80% as seen in Fig 3 B). Also, Images of the cells cultured with 200 ppm of MoS$_2$ NR-Au NPs are shown in Fig. 3 C (c). No dead cell can be seen in Fig. 3 C (c) also, which indicates the biocompatibility of MoS$_2$ NR-Au NPs. Fig. 3 C (b) is after NIR irradiation without MoS$_2$ NR-Au NPs. After trypan blue staining, any dead cell (blue stained) can be seen hardly either in Fig 3 C (a) or Fig 3 C (b). This indicates that 808 nm NIR laser itself do not help in elevating the temperature of the cell medium directly.

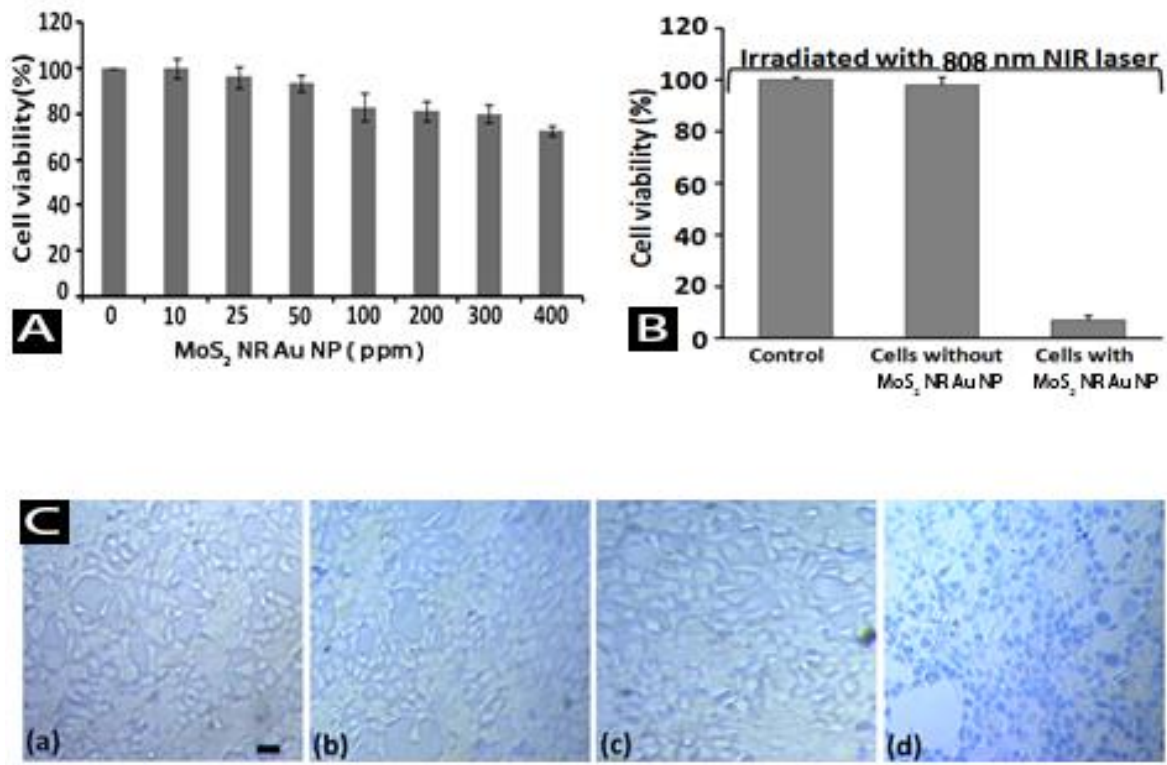

Figure 3: A) Cytotoxicity of $MoS_2$ NR Au NPs system. B) Quantitative Viability of cells irradiated C) Optical images of in-vitro NIR photothermal ablation of SiHa cancerous cells. a) Control healthy cells without laser irradiation in absence of $MoS_2$ NR-Au NPs. b) Cell viability 24 hrs post laser irradiation c) SiHa cells cultured with 200 ppm of $MoS_2$ NR-Au NPs without irradiation d) SiHa cancerous cells cultured with 200ppm $MoS_2$ NR-Au NPs followed by laser irradiation.

By the way of conclusion, although the photothothermal destruction of cancer cell have been reported by both gold nanoparticles[9] and recently by MoS2 nanosheets[23,26], the present work integrates the excellent photothermal properties of both gold nanoparticles and MoS2 nanoribbons to take out the benefit of synergistically improved photothermal effect. The synthesized $MoS_2$ NR-Au NPs not only present the better biocompatibility (even till concentrations as high as 300ppm) but brings about outstanding photothermal effects with a biocompatible and low concentration of 200 ppm towards the complete killing of cancer cells at low laser power (808nm) and less time of irradiation (3 minutes). Further the proposed synthesis route involves facile route for bulk production of $MoS_2$ NR and the quick microwave based decoration of Au NP on them. With such parameters the present work propels an accelerated research on this domain in the coming future aiming on more scrutinized experiments exploiting the strong mechanical bonding ability of these nanoribbons (which is a function of its size and aspect ratio) which can possibly be used to hold the drugs onto it and finally selectively hold the target more firmly for much enhanced combined photo and chemo therapy. Also the use of metallic (1T) $MoS_2$ matrix instead of the semiconducting (H) $MoS_2$ NR matrix used in the present work is yet another exciting domain to look for after acknowledging the potential of presently proposed hybrid system. Along with above mentioned scopes our future work would involve efforts for a mechanistic study of this enhanced effect as well. Realizing these potentials the present system definitely serves as an archetype for various applications requiring heat transduction and biocompatibility at the same time; among which the efficacy for photothermal ablation of cancerous cells has been attested here.

# Gold Nanoparticles supported on MoS$_2$ Nanoribbons matrix as a biocompatible and water dispersible platform for enhanced photothermal ablation of cancerous cells using harmless near infrared irradiation.

*Shobhit Pandey$^{a,\#}$, Anushka Bansal$^{a,\#}$, Vinod Kumar$^b$, Himanshu Mishra$^c$, Rohit Kumar Agrawal$^a$, Sunayana Kashyap $^d$, Bratindranath Mukherjee$^a$, Preeti Saxena$^d$, and Anchal Srivastava$^{c*}$*

$^a$ *Department of Metallurgical Engineering, Indian Institute of Technology (Banaras Hindu University), Varanasi-221005, India*

$^b$ *Institute of Nano Science and Technology, Habitat Centre, Phase- 10, Sector- 64, Mohali, - 160062 India*

$^c$ *Department of Physics, Banaras Hindu University (BHU), Varanasi-221005, India*

$^d$ *Department of Zoology, Banaras Hindu University, Varanasi, 221005, India*

*# Equally contributing authors.*

**Address correspondence to:**

E-mail: anchalbhu@gmail.com (Dr. Anchal Srivastava)

Phone No.: +91-9453203122, Fax: +91-542 2368174

## S1. Synthesis of the MoS$_2$ NR- Au NPs system.

### a) Synthesis of h-MoO$_3$ nanorods (MoO$_3$ nr)

As a first step of the synthesis process, 1.2g of ammonium heptamolybdate tetra hydrate (AHM, (NH$_4$)$_6$Mo$_7$O$_{24}$.4H$_2$O) was dissolved in a mixed solution having 10% HNO$_3$ and DW. After the dissolution was complete this solution was put in a Teflon-lined stainless steel autoclave (20 ml capacity) and heated at 180°C for 12 hours, following which it was cooled under normal conditions. Finally, the light grey precipitate was obtained by centrifugation, which was washed thoroughly with DW and ethanol several times before drying at 40°C for 12 hours.

### b) Preparation of MoS$_2$ Nanoribbons (MoS$_2$ NR)

The above obtained MoO$_3$ nr through optimized process, was then loaded in a ceramic boat in a tube furnace. Inert gas (Argon in our case) & H$_2$S atmosphere was provided following a temperature profile in which the temperature of the furnace was first slowly raised at 20$^0$C per minute to 800$^0$C; at which it was held for 1hr and then furnace cooled. This resulted in the facile synthesis of MoS$_2$ NR. The pressure of the gas passed was maintained at 5 torr**,** which was suitable for efficient conversion of these MoO$_3$ nr to MoS$_2$ NR. The temperature range was chosen based on earlier reports. [1]

### c) Decoration of Gold NPs on MoS$_2$ NR

Above synthesized MoS$_2$ NR was made in Ethylene glycol with 1% doping of Gold using Chloroauric Acid. Microwave reduction of this solution, was then performed for 15 Seconds for the decoration of Gold Nanoparticles (Au NPs) in MoS$_2$.[2] The powder of this Au NPs decorated MoS$_2$ NR was obtained from the solution by Centrifuging. This powder was then dried and used for further experimental work and characterization techniques. From the SEM micrographs, it is evident that the gold nanoparticles have been decorated on these MoS$_2$ Nanoribbons.

### S2. MoO$_3$ Characterizations

#### a) XRD measurements

Figure S1 shows the XRD pattern of the synthesized samples. All the identified peaks of the synthesized samples are indexed for h-MoO$_3$ (JCPDS card no. 21-0569, space group P63/m, a = 10.53 Å and c = 14.98 Å). The high intensity of diffraction peaks of the planes (210), (220), (410), (610), (110) etc. represents an anisotropic growth. Peaks due to the other phases are not observed indicating the high purity of h-MoO$_3$.

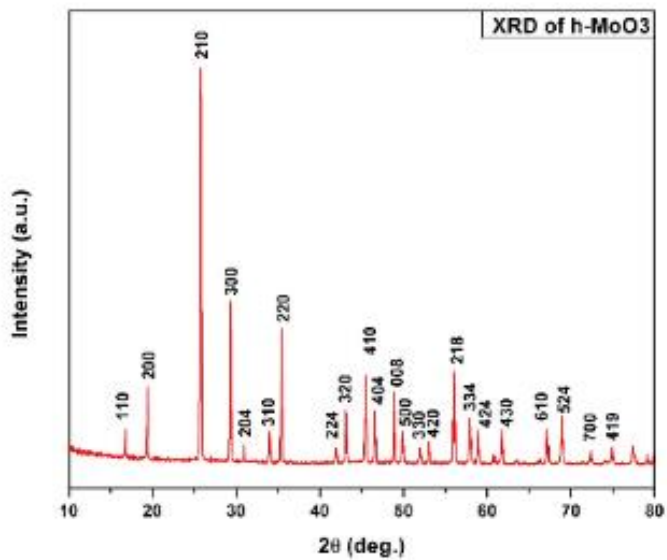

*Figure S 1: XRD of MoO$_3$ Nanorods*

### S3. MoS$_2$ NR characterization:

#### a) XRD Analysis.

XRD was used to characterize the crystal structure and phase purity of the samples (Fig S 2). It can be observed that the diffraction peaks were nearly consistent with the reported value of MoS$_2$ from JCPDS no. (77-1716), indicating a successful synthesis of 2H-MoS$_2$ nano ribbon. The corresponding indexing of the peaks resulted in the (002), (100), (103), (110) and (112) planes

with the most prominent peak coming around the (002) plane. This is in agreement with the earlier reports.[3]

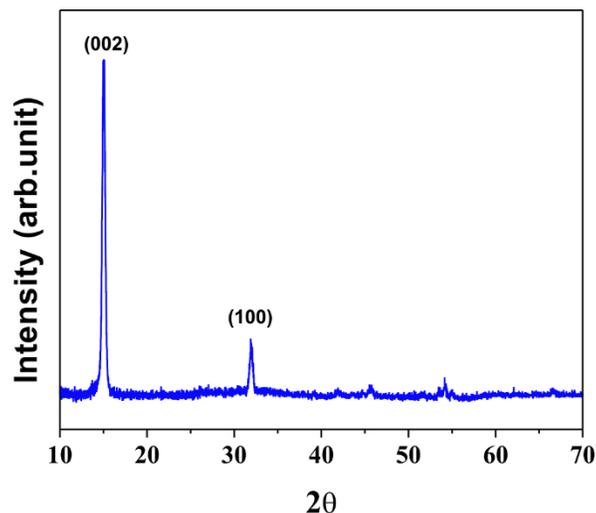

*Figure S 2: XRD pattern of MoS₂ Nanoribbons*

**b) TEM analysis**

The TEM images in Fig S4, shows the detailed structure of $MoS_2$ nano ribbons and its bundle like structure. Its thinness is confirmed by its electron transparency and uniformity over the width. The large no. of concentric rings in the electron diffraction pattern suggests that the $MoS_2$ formed is crystalline in nature. Further the slight broadening and streaking/ arcing of the spots can be seen which is in agreement with the ultra- thin nano-ribbon morphology. The rings have been perfectly indexed to the (100) (103) and (105) crystal diffractions of the hexagonal $MoS_2$ confirming its identity. Further this $MoS_2$ Nanoribbon matrix has been used for the decoration of gold nanoparticles.

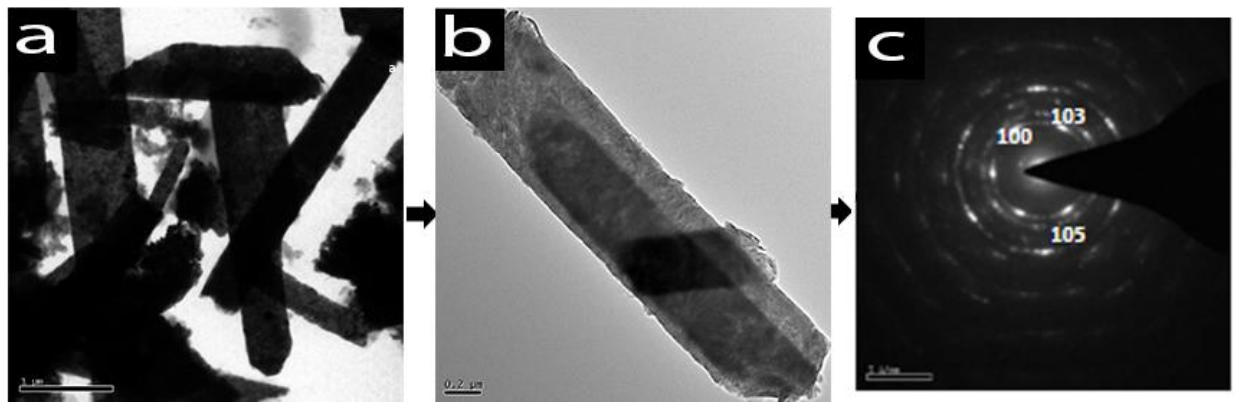

*Figure S 3: TEM micrograph of MoS₂ NR matrix. a) Dispersion of MoS₂ NR in DW. b) A Single MoS₂ NR showing good uniformity in thickness. c) Indexing of the corresponding diffraction pattern confirming the planes of MoS₂.*